\documentclass[12pt]{article}

\begin{document}
\begin{titlepage}
\title{ Science as a Culture - Its Implications \footnote{Inaugural Lecture 
at `CMDAYS 99' (held at Jadavpur University, Kolkata, India) 
by Prof. Shyamal Sengupta}}
\author{}
\date{
Shyamal Sengupta}
\maketitle
\abstract{This is a lecture on the ethics and role of science 
in promoting rational and objective thinking in society. 
It was delivered by Prof. 
Shyamal Sengupta of Kolkata, India. Prof. Sengupta, who passed away recently,
has inspired generations of Indian physicists by his rational viewpoints
on science and by teaching his students the importance of a scientific outlook.
We, some of his former students dedicate this small corner of the archive 
to his fond memory. \footnote{Any comments on this article can be emailed to
Supratim Sengupta (sengupta@physics.mcmaster.ca), 
Tapobrata Sarkar (tapo@ictp.trieste.it), 
Subrata Bal (subrata@gauge.scphys.kyoto-u.ac.jp),
Saurya Das (saurya.das@uleth.ca) }}
\end{titlepage}

\noindent
 Science as a methodology for probing the secrets of nature is
 well known. During the last four hundred years science has created a gigantic
 store house of exact knowledge of the physical world. On the other hand science
 as a technology has maximised the use of scientific knowledge to transform the
 economic conduction of the society.

 There is a third aspect of science which is mostly overlooked and rarely
 discussed. This is, science as a culture. It has deep and important
 implications, though not as clearly manifested as the other two. In India, ours
 is a scattered and mostly ineffective scientific community. So far our community
 needs are, to a large extent, met by the Western society. I believe one reason
 for this is, we are lacking, collectively speaking, in scientific culture.\\

\noindent
 {\bf Science as a Culture} : In 1874 Francis Galton, the father of biometry, 
 carried out a survey of 180 top ranking scientists of England and Wales. The 
 result was published in a book entitled English Men of Science. Their Nature 
 and Nurture. Some of the findings on the religious belief of scientists were 
 as follows.

 \begin{itemize}
 \item
 Eight out of ten scientist were members of the conventional church of
 England.
 \item
 Most of them showed a concern for human welfare much above the average
 people.
 \item
 Even these, who categorically stated, they had a religious bend,
 did not believe in the christian doctrine of original sin and the doctrine of
 life after death.
 \end{itemize}

     A typical answer was {\it ``though I believe in Christian religion, I 
 do not accept
 the authority of the Church to teach me how I can attain emancipation. I strongly
 deny any interference with my freedom of thought and the freedom of my
 conscience.''} It was clear, the study of science was changing the outlook of
 scientists in matters not directly connected with science. This we call 
 the {\it \bf{cultural
 aspects of science}}. The study of science affects a scientist's thinking on life
 in general, his judgements, his value system. However, such an effect seems
 possible only if the pursuit of science is inspired at least partly, by the
 primary motivation. Today, particularly in developing countries like India, many
 persons take to science because of money, fame, power or for some other
 secondary reasons. In such cases it is possible to develop a compartmentalised
 approach, not allowing scientific thinking to percolate into one's religious or
 cultural beliefs. A scientific community can tolerate only a limited number of
 such workers. If the number is large, the community ceases to be effective.
 
 The primary motivation for science starts from a natural curiosity about the
 objective world and if properly nurtured, turns into a passion for problem
 solving. Inability to arrive at a solution is sometimes so discomforting that if
 some one else gives the solution, one feels relieved and delighted, even though
 one misses the credit. When one really enjoys scientific activity, the solution 
 of a problem becomes of primary importance, who solves the problem is relegated to
 the second place. One need not be a selfless idealist to achieve this cultural
 state of the mind. It seems natural when one is working in the right atmosphere
 of a scientific community where the primary emphasis is on knowledge rather than
 on the individual. Those who contribute to scientific knowledge are important,
 but the totality of this knowledge is far more significant. As Edwin Hubble
 remarked in his book ``The Realm of the Nebulae,''  -  {\it ``Today the least of the
 men of science commands a wider prospect (than Newton). Even the giants are
 dwarfed by the great edifice in which their achievements are incorporated.''}

 The emergence of such an objective mind is an important contribution of science
 to human culture. To such a mind, facts are sacred and statements based on facts
 are the only ones worthy of examination. Accurate and reproducible facts are the
 basic elements on which the grand cumulative edifice of scientific knowledge
 stands. Conceptual and theoretical notions about the objective world undergo
 abrupt changes as theories advance. But facts do not change, they are made more
 and more accurate. Alteration, distortion, or selective choice of facts to fit a
 particular theory or idea is highly unethical in the culture of science.

 The habit of objective thinking has overtones beyond the realm of science. For
 example, it offers a new perspective to look at outstanding scientists for
 personalities in other spheres such that their image can never be moulded in
 pattern with the spiritual concept of an infallible guru. A successful scientist
 is highly respected for his achievement, but this does not prevent the community
 to reject or criticize his other ideas which are not supported by facts.
 Galileo's brilliant studies in mechanics does not justify his theory that tides
 are caused by the earth's rotation. Success of Newton's gravitational theory does
 not justify his corpuscular theory of light. Nobody is infallible and authority
 cannot be sited as proof. Indeed, these ideas mark a big step forward in the
 domain of human culture.

 Coming to the group of talented scientific workers who constitute the core of
 the scientific community, objective attitude makes at possible for two or more
 persons to collaborate in solving a problem, though they may not like each other
 because of differences in temperament. This is possible only when all the
 collaborators agree that the solution of the problem is urgent and important.
 Without such collaborative efforts a meaningful scientific community cannot be
 created and it would have been impossible for science to make the gigantic
 progress we see now. The science community acting as a wonderful sieve has
 allowed only the objective contributions from individual scientists to enter the
 palace of scientific knowledge, leaving out their oddities and subjective
 inclinations. When we think of Kepler constructing a horoscope of the cosmos, 
 Newton spending a huge lot of time on theology and Einstein preoccupied with
 pure mathematical constructions of the mind as gateway to understanding nature,
 then only can we appreciate the true significance of the important role played
 by the science community.

 We have mentioned about ethics in science in connection with importance of
 facts. As science was growing slowly in Europe, a well defined ethical standard
 was also emerging. This standard is not implemented by law, indeed there is no
 codified statement about the ethical principles to be observed. These evolved as
 conventions and were implemented through community reactions. An unethical referee   
 may reject a good paper and steal the idea to write a paper himself. But the
 community reaction would be so strong that few venture to take the risk. A few
 years back Nature published a paper, reporting highly doubtful experimental
 results, on `water has memory'. The scientific community reacted so strongly
 that the editor had to institute an enquiry committee and ultimately he
 expressed regret and declared - ``regard the paper as unpublished.'' Similar
 reactions were generated when supposed results of cold fusion experiments were
 first declared, not in a paper, but in a press conference. Weakness of our
 scientific community is shown by the fact that departure from ethics 
 in India is quite frequent and go mostly unpunished. This is why some years ago
 a society for scientific values was set up by Professor A.S. Paintal. We do not
 go into further details. It is hoped that from the brief discussion given above
 the reader can develop a broad understanding of the nature of scientific
 culture.\\

 \noindent
 {\bf Implications :} Our assertion is not that science is such a 
 distinct culture that
 successful studies of science cannot be carried out in the background of the
 culture of our society. In fact science has spread beyond the realm of Western
 culture, as in Japan. The point is, essential elements of scientific culture
 have to be integrated within the existing culture for developing an effective
 scientific community in the country. May be, in the long run, interaction
 between the two would modify both to some extent.

 1900 to 1930 is the so called golden age of physics in India. A few Indian
 physicists, starting with Professor J.C. Bose, achieved remarkable success and
 received international acclaim.

 Considering the Indian background, the whole phenomenon seemed a miracle. The
 dazzling event boosted our national pride and deeply impressed young Indian
 physicists of subsequent generations. Unfortunately there was also a different
 fall out. Instead of interpreting the event as exceptional, we developed the
 notion that this is the normal way discoveries in science are made . As
 Professor Chandrasekhar puts it, {\it ``..... it gave a false picture, that making
 discoveries is easy. It had a distorting effect. I mean it was alright in the
 twenties, but extended into later period, it was not. I started with a totally
 glamorous view of science that persisted so long as I was in India. But going to
 England was a shattering experience.''}

 The normal and continuous development of scientific knowledge occurs through the
 efforts of a competent and dedicated community of scientists. It is in the
 background of such a community alone that a genius can flower into full glory.
 Unfortunately, neither the scientists of the golden era nor any subsequent group
 gave much though to this problem. Today we have so many people doing physics in
 our country, but to most of them it is an entirely individualistic pursuit. We
 miss the sense of adventure in our work. Instead there is isolation and
 frustration. This is well reflected in a letter Professor K.S. Krishnan wrote to
 Chandrasekhar in 1936, on the question of latter's return to India from USA.

 {\it ``On this side at any rate, I know the condition and with all my optimism and
 enthusiasm for science. I feel the present scientific atmosphere quite
 oppressive and it may be an advantage if you postpone your return to India by
 one or two years.''}

 Andre Weil, an eminent French mathematician of this century, came to India at a
 very young age and joined Aligarh Musilim University as head of the department
 of mathematics in 1930 and stayed for two years. He also visited Dacca
 University and met Professor S. N. Bose there. In a recollection in 1979 (recorded
 by Dr. R. Banerjee), Professor Weil commented - {\it ``The early thirties in India 
 was a disturbed period, due to non-coperation movement. Many of us hoped that the
 upheaval, initially with a political content, could be an occasion for the
 national leaders for creating a condition for the flourishment of new ideas and
 foundation of scientific studies and research in the country. I am sorry to say
 that this expectation has not been fulfilled. As I found then, the atmosphere of
 intrigue that reigned in the universities was paralysing in every sense. I am
 now retired as a professor, and in my long career, I have seen many examples of
 intrigues and personal rivalry in University sphere in various continents. But,
 believe me, what I have witnessed during two years in India exceed by far what I
 have seen during the rest of my life.''} [We quote Weil's comments at some length
 because he developed a serious interest in India and its culture. He met, among
 many Indian leaders, both Tagore and Gandhi. He learnt Sanskrit to read the Gita 
 and the Mahabharata in the original. This comment made in connection with his
 reminiscences of Professor S. N. Bose is as yet unpublished. I thank Dr.Purnima
 Sinha, a student of Prof. Bose for sending me a copy of the interview].

 Has the atmosphere changed radically after independence ? There is hardly any
 indication. When Professor Chandrasekhar was requested to be the chairman of
 the Atomic Energy Commission after Bhabha's death, he came to India and visited
 several scientific Institutions and also met many scientists. Ultimately he
 declined the offer and before leaving India commented, {\it ``the problem is not 
 with money, it is the atmosphere which is discouraging.''}

 In 1985, in an address before the IPA, Professor Virendra Singh, the 
 chairman, said, {\it ``presently the physics community in the country is rather
 fragmented. A feeling of isolation is felt by most of the members.''}

 The above discussion reveals the true significance of science as a culture. We
 cannot create the right atmosphere for basic scientific activities, both
 teaching and research unless we have imbibed scientific culture, i.e. unless our
 minds are genuinely involved with the problem of the objective world, the
 problems which we discuss either in the class rooms or in research seminars. How
 this transformation can be achieved is a different and complex problem. It is
 not within the scope of the present discussion.

 We conclude by mentioning another important implication of scientific culture.
 Most of the existing cultures are based on religion and they originated at a
 time when little exact knowledge of the physical world existed. After the
 spectacular advance of modern science it has become more and more difficult to
 accept these elements; of a culture which are bases on objective falsehood. It
 is not surprising, this impact is strongest on the christian culture. A few
 years back, Galileo's case was reopened by the Pope and the mistake in the
 judgement was admitted. In the case of the shroud of Turin, a shroud believed 
 to be used to cover the body of Jesus Christ after it was removed from the cross,
 scientists in 1978 wanted a small piece of the shroud to do carbon dating to
 establish its age. The cardinal of the church refused. But in 1988, the
 cardinal, of his own accord, sent pieces of the shroud to three laboratories.
 After testing, it was definitely established that the date of origin was later
 than 1200 AD, most probably between 1260 to 1390 AD. The myth was exploded. Some
 years ago, the newly appointed Archbishop in England declared that a person can
 remain a devout christian without believing in any of the numerous miracles
 described in the New Testament. Initially there were angry protests with a
 demand for the removal of the Archbishop. But nothing serious happened. Very
 recently, on 23rd July 1999, Pope Paul redefined the concept of heaven: it is not
 a place above the clouds, it is a state of being close to God, as hell is a
 state of being separated from God. It seems that objective truth creates a
 psychological pressure on all thinking men which can be relieved only by getting
 rid of the objectively false beliefs. The process is definitely very slow at
 present. But may be at some critical stage it would be much faster.

 Some day in the distant future, we hope all cultures of the world would free
 themselves of elements of faith which contradict the objective knowledge of
 science. Then all cultures will come closer and communication across the
 cultural barriers will be easier. The age old conflict between the cultures may
 then be resolved with much less difficulty. Perhaps, this would be the greatest
 contribution of science to humanity .
\end{document}